\documentclass[preprintnumbers,amsmath,amssymb,11pt,floatfix,superscriptaddress,nofootinbib]{revtex4}
\usepackage{graphicx}
\usepackage{epsfig}
\usepackage{bm}
\usepackage{amsfonts}

\newcommand{\la}{\lambda}

\def\be {\begin{equation}}
\def\ee {\end{equation}}
\def\ba {\begin{eqnarray}}
\def\ea {\end{eqnarray}}
\newcommand{\bear}{\begin{eqnarray}}

\newcommand{\eear}{\end{eqnarray}}

\newbox\pippobox

\def\6{\partial}

\def\la{\Lambda}
\def\a{\alpha}

\def\sq
\def\a{\alpha}

\begin{document}
\title{New agegraphic dark energy
in Ho\v{r}ava-Lifshitz cosmology}

\author{ Mubasher Jamil} \email{mjamil@camp.edu.pk}
\affiliation{Center for Advanced Mathematics and Physics,\\
National University of Sciences and Technology, H-12, Islamabad,
Pakistan}

\author{Emmanuel N. Saridakis} \email{msaridak@phys.uoa.gr}
 \affiliation{College of Mathematics
and Physics,\\ Chongqing University of Posts and Telecommunications,
Chongqing, 400065, P.R. China }

\begin{abstract}
We investigate the new agegraphic dark energy scenario in a
universe governed by Ho\v{r}ava-Lifshitz gravity. We consider both
the detailed and non-detailed balanced version of the theory, we
impose an arbitrary curvature, and we allow for an interaction
between the matter and dark energy sectors. Extracting the
differential equation for the evolution of the dark energy density
parameter and performing an expansion of the dark energy
equation-of-state parameter, we calculate its present and its
low-redshift value as functions of the dark energy and curvature
density parameters at present, of the Ho\v{r}ava-Lifshitz running
parameter $\lambda$, of the new agegraphic dark energy parameter
$n$, and of the interaction coupling $b$. We find that
$w_0=-0.82^{+0.08}_{-0.08}$ and $w_1=0.08^{+0.09}_{-0.07}$.
Although this analysis indicates that the scenario can be
compatible with observations, it does not enlighten the discussion
about the possible conceptual and theoretical problems of
Ho\v{r}ava-Lifshitz gravity.
\end{abstract}

\maketitle
\newpage
\section{Introduction}

Many recent cosmological observations, such as SNIa \cite{1}, WMAP
\cite{2}, SDSS \cite{3} and X-ray \cite{4} support the idea that
the universe is experiencing an accelerated expansion. A first
direction that could provide an explanation of this remarkable
phenomenon is to introduce the concept of dark energy, with the
most obvious theoretical candidate being the cosmological
constant. However, at least in an effective level, the dynamical
nature of dark energy can also originate from   a variable
cosmological ``constant'' \cite{varcc}, or from various fields,
such is a canonical scalar field (quintessence) \cite{quint}, a
phantom field \cite{phant}, or the combination of quintessence and
phantom in a unified model named quintom \cite{quintom}. The
second direction that could explain the acceleration is to modify
the gravitational theory itself, such in the generalization to
$f(R)$-gravity \cite{12}, to scalar-tensor theories with
non-minimal coupling \cite{9} etc.

Going beyond the aforementioned effective description requires a
deeper understanding of the underlying theory of quantum gravity,
unknown at present. However, physicists can still make some
attempts to probe the nature of dark energy according to some
basic quantum gravitational principles. An interesting such an
attempt is the so-called ``Holographic Dark Energy'' proposal
\cite{Hsu:2004ri,Li:2004rb}. Its framework is the black hole
thermodynamics  and the connection (known from AdS/CFT
correspondence) of the UV cut-of of a quantum field theory, which
gives rise to the vacuum energy, with the largest distance of the
theory \cite{Cohen:1998zx}. Thus, determining an appropriate
quantity $L$ to serve as an IR cut-off, imposing the constraint
that the total vacuum energy in the corresponding maximum volume
must not be greater than the mass of a black hole of the same
size, and saturating the inequality, one identifies the acquired
vacuum energy as holographic dark energy: $
\rho_\Lambda=\frac{3c^2}{8\pi G L^2}, $ where $G$ is the
gravitational Newton's
 constant and $c$ a constant. The holographic dark
energy scenario has been tested and constrained by various
astronomical observations \cite{obs3a} and it has been extended to
various frameworks \cite{nonflat,holoext,intde}.

A specific application of holographic dark energy is obtained when
the age of the universe $T=\int dt$ is used as the IR cut-off $L$,
the so-called agegraphic dark energy scenario \cite{Cai:2007us}.
However, since this scenario cannot describe consistently the
matter-dominated period, it was extended to the new agegraphic
dark energy, namely under the use of the conformal time $\eta$ as
the IR cut-off $L$ \cite{Wei:2007ty,Wu:2008jt}.

On the other hand, concerning the gravitational background of the
universe, almost one year ago  Ho\v{r}ava proposed a
power-counting renormalizable theory with consistent ultra-violet
(UV) behavior \cite{hor1}. Although presenting an infrared (IR)
fixed point, namely General Relativity, in the  UV the theory
exhibits an anisotropic, Lifshitz scaling between time and space.
Ho\v{r}ava-Lifshitz gravity has been studied and extended in
detail \cite{Cai:2009dx}
 and it has been applied as the cosmological framework of the universe
\cite{Calcagni:2009ar,Lu:2009em,Sotiriou:2009bx,Wu:2009ah}.

In the present work we are interested in investigating the new
agegraphic dark energy scenario in a universe governed by
Ho\v{r}ava-Lifshitz gravity. The plan of the paper is the
following: In  section \ref{model} we present Ho\v{r}ava-Lifshitz
cosmology and in section \ref{NADEsec} we analyze the new
agegraphic dark energy scenario. In section \ref{NADEHL} we
construct the scenario of new agegraphic dark energy in
Ho\v{r}ava-Lifshitz cosmology, both in the simple as well as in
the interacting form, extracting the differential equations that
determine the evolution of the dark energy density parameter. In
section \ref{cosmimpl} we discuss the cosmological implications of
such a scenario, and in particular we calculate the values and the
bounds of the dark-energy equation-of-state  parameter assuming a
linear low-redshift parametrization. Finally, in section
\ref{Conclusions} we summarize the obtained results.

\section{Ho\v{r}ava-Lifshitz cosmology}
\label{model}

In this section we briefly review the scenario where the
cosmological evolution is governed by Ho\v{r}ava-Lifshitz gravity
\cite{Calcagni:2009ar}. The dynamical variables are the lapse and
shift functions, $N$ and $N_i$ respectively, and the spatial
metric $g_{ij}$ (roman letters indicate spatial indices). In terms
of these fields the full metric is written as:
\begin{eqnarray}
ds^2 = - N^2 dt^2 + g_{ij} (dx^i + N^i dt ) ( dx^j + N^j dt ) ,
\end{eqnarray}
where indices are raised and lowered using $g_{ij}$. The scaling
transformation of the coordinates reads: $
 t \rightarrow l^3 t~~~{\rm and}\ \ x^i \rightarrow l x^i
$.

\subsection{Detailed Balance}

The gravitational action is decomposed into a kinetic and a
potential part as $S_g = \int dt d^3x \sqrt{g} N ({\cal L}_K+{\cal
L}_V)$. The assumption of detailed balance \cite{hor1}
  reduces the possible terms in the Lagrangian, and it allows
for a quantum inheritance principle, since the $(D+1)$-dimensional
theory acquires the renormalization properties of the
$D$-dimensional one. Under the detailed balance condition
 the full action of Ho\v{r}ava-Lifshitz gravity is given by
\begin{eqnarray}
 S_g =  \int dt d^3x \sqrt{g} N \left\{
\frac{2}{\kappa^2} (K_{ij}K^{ij} - \lambda K^2) +\frac{\kappa^2}{2
w^4} C_{ij}C^{ij}
 -\frac{\kappa^2 \mu}{2 w^2}
\frac{\epsilon^{ijk}}{\sqrt{g}} R_{il} \nabla_j R^l_k +
\frac{\kappa^2 \mu^2}{8} R_{ij} R^{ij}
     \right. \nonumber \\
+\left.    \frac{\kappa^2 \mu^2}{8( 3 \lambda-1)} \left[ \frac{1 -
4 \lambda}{4} R^2 + \Lambda  R - 3 \Lambda ^2 \right] \right\},
\label{acct}
\end{eqnarray}
where
\begin{eqnarray}
K_{ij} = \frac{1}{2N} \left( {\dot{g}_{ij}} - \nabla_i N_j -
\nabla_j N_i \right)
\end{eqnarray}
is the extrinsic curvature and
\begin{eqnarray} C^{ij} \, = \, \frac{\epsilon^{ijk}}{\sqrt{g}} \nabla_k
\bigl( R^j_i - \frac{1}{4} R \delta^j_i \bigr)
\end{eqnarray}
the Cotton tensor, and the covariant derivatives are defined with
respect to the spatial metric $g_{ij}$. $\epsilon^{ijk}$ is the
totally antisymmetric unit tensor, $\lambda$ is a dimensionless
constant and the variables $\kappa$, $w$ and $\mu$ are constants
with mass dimensions $-1$, $0$ and $1$, respectively. Finally, we
mention that in action (\ref{acct}) we have already performed the
usual analytic continuation of the parameters $\mu$ and $w$ of the
original version of Ho\v{r}ava-Lifshitz gravity, since such a
procedure is required in order to obtain a realistic cosmology
\cite{Lu:2009em} (although it could fatally affect the
gravitational theory itself). Therefore, in the present work
$\Lambda $ is a positive constant, which as usual is related to
the cosmological constant in the IR limit.

Lastly, in order to incorporate the (dark plus baryonic) matter
component one adds a cosmological stress-energy tensor to the
gravitational field equations, by demanding to recover the usual
general relativity formulation in the low-energy limit
\cite{Sotiriou:2009bx}. Thus, this matter-tensor is a
hydrodynamical approximation with its energy density $\rho_M$ and
pressure $p_M$ (or $\rho_M$ and its equation-of-state parameter
$w_M\equiv p_M/\rho_M$) as parameters.

Now, in order to focus on cosmological frameworks, we impose the
so called projectability condition \cite{Calcagni:2009ar} and use
a Friedmann-Robertson-Walker (FRW)  metric,
\begin{eqnarray}
N=1~,~~g_{ij}=a^2(t)\gamma_{ij}~,~~N^i=0~,
\end{eqnarray}
with
\begin{eqnarray}
\gamma_{ij}dx^idx^j=\frac{dr^2}{1- k r^2}+r^2d\Omega_2^2~,
\end{eqnarray}
where $ k=-1,0,+1$ corresponding  to open, flat, and closed
universe respectively. By varying $N$ and $g_{ij}$, we obtain the
equations of motion:
\begin{eqnarray}\label{Fr1fluid}
H^2 = \frac{\kappa^2}{6(3\la-1)} \rho_M
+\frac{\kappa^2}{6(3\la-1)}\left[ \frac{3\kappa^2\mu^2
k^2}{8(3\lambda-1)a^4} +\frac{3\kappa^2\mu^2\Lambda
^2}{8(3\lambda-1)}
 \right]-\frac{\kappa^4\mu^2\Lambda  k}{8(3\lambda-1)^2a^2}
\end{eqnarray}
\begin{eqnarray}\label{Fr2fluid}
\dot{H}+\frac{3}{2}H^2 = -\frac{\kappa^2}{4(3\la-1)} w_M\rho_M
-\frac{\kappa^2}{4(3\la-1)}\left[\frac{\kappa^2\mu^2
k^2}{8(3\lambda-1)a^4} -\frac{3\kappa^2\mu^2\Lambda
^2}{8(3\lambda-1)}
 \right]-\frac{\kappa^4\mu^2\Lambda  k}{16(3\lambda-1)^2a^2} ,
\end{eqnarray}
where we have defined the Hubble parameter as $H\equiv\frac{\dot
a}{a}$. As usual, $\rho_M$ follows the standard evolution equation
\begin{eqnarray}\label{rhodotfluid}
&&\dot{\rho}_M+3H(1+w_M)\rho_M=0.
\end{eqnarray}

Observing the above Friedmann equations, concerning the
dark-energy sector we can define
\begin{equation}\label{rhoDE}
\rho_{DE}\equiv \frac{3\kappa^2\mu^2 K^2}{8(3\lambda-1)a^4}
+\frac{3\kappa^2\mu^2\Lambda ^2}{8(3\lambda-1)}
\end{equation}
\begin{equation}
\label{pDE} p_{DE}\equiv \frac{\kappa^2\mu^2
K^2}{8(3\lambda-1)a^4} -\frac{3\kappa^2\mu^2\Lambda
^2}{8(3\lambda-1)}.
\end{equation}
The term proportional to $a^{-4}$ is the usual ``dark radiation
term'', present in Ho\v{r}ava-Lifshitz cosmology
\cite{Calcagni:2009ar}, while the constant term is just the
explicit cosmological constant. Therefore, in expressions
(\ref{rhoDE}),(\ref{pDE}) we have defined the energy density and
pressure for the effective dark energy, which incorporates the
aforementioned contributions. Finally, note that using
(\ref{rhoDE}),(\ref{pDE}) it is straightforward to show that these
 dark energy quantities satisfy the
standard evolution equation:
\begin{eqnarray}
\label{DEevol} &&\dot{\rho}_{DE}+3H(\rho_{DE}+p_{DE})=0.
\end{eqnarray}

As a last step, requiring these expressions to coincide with the
standard Friedmann equations, in units where $c=1$  we set
\cite{Calcagni:2009ar}:
\begin{eqnarray}
&&G_{\rm cosmo}=\frac{\kappa^2}{16\pi(3\lambda-1)}\label{simpleconstants0a}\\
&&\frac{\kappa^4\mu^2\Lambda}{8(3\lambda-1)^2}=1,
\label{simpleconstants0}
\end{eqnarray}
where $G_{\rm cosmo}$ is the ``cosmological'' Newton's constant.
We mention that in theories with Lorentz invariance breaking (such
is Ho\v{r}ava-Lifshitz one) the ``gravitational'' Newton's
constant $G_{\rm grav}$, that is the one that is present in the
gravitational action, does not coincide with the ``cosmological''
Newton's constant $G_{\rm cosmo}$, that is the one that is present
in Friedmann equations, unless Lorentz invariance is restored
\cite{Carroll:2004ai}. For completeness we mention that in our
case
\begin{eqnarray}
G_{\rm grav}=\frac{\kappa^2}{32\pi}\label{Ggrav},
\end{eqnarray}
as it can be straightforwardly read from the action (\ref{acct}).
Thus, it becomes obvious that in the IR ($\lambda=1$), where
Lorentz invariance is restored, $G_{\rm cosmo}$ and $G_{\rm grav}$
coincide.

Using the above identifications, we can re-write the Friedmann
equations (\ref{Fr1fluid}),(\ref{Fr2fluid}) as
\begin{equation}
\label{Fr1b} H^2+\frac{k}{a^2} = \frac{8\pi G_{\rm
cosmo}}{3}\left(\rho_M+\rho_{DE}\right)
\end{equation}
\begin{equation}
\label{Fr2b} \dot{H}+\frac{3}{2}H^2+\frac{k}{2a^2} = - 4\pi G_{\rm
cosmo}\left(w_M\rho_M+w_{DE}\rho_{DE}\right),
\end{equation}
where we have introduced the effective dark energy
equation-of-state parameter $ w_{DE}\equiv p_{DE}/\rho_{DE}$.

\subsection{Beyond Detailed Balance}

The above formulation of Ho\v{r}ava-Lifshitz cosmology has been
performed under the imposition of the detailed-balance condition.
However, in the literature there is a discussion whether this
condition leads to reliable results or if it is able to reveal the
full information of Ho\v{r}ava-Lifshitz
 gravity \cite{Calcagni:2009ar}. Thus, one
 should study also the Friedmann equations in the case
 where detailed balance is relaxed. In such a case one can in
 general write
 \cite{Sotiriou:2009bx}:
\begin{eqnarray}\label{Fr1c}
H^2 = \frac{2\sigma_0}{(3\la-1)} \rho_M +\frac{2}{(3\la-1)}\left[
\frac{\sigma_1}{6}+\frac{\sigma_3 K^2}{6a^4} +\frac{\sigma_4
K}{6a^6}
 \right]+\frac{\sigma_2}{3(3\la-1)}\frac{ K}{a^2}
\end{eqnarray}
\begin{eqnarray}\label{Fr2c}
\dot{H}+\frac{3}{2}H^2 = -\frac{3\sigma_0}{(3\la-1)} w_M\rho_M
-\frac{3}{(3\la-1)}\left[ -\frac{\sigma_1}{6}+\frac{\sigma_3
K^2}{18a^4} +\frac{\sigma_4 K}{6a^6}
 \right]+
 \frac{\sigma_2}{6(3\la-1)}\frac{ K}{a^2},
\end{eqnarray}
where $\sigma_0\equiv \kappa^2/12$, and the constants $\sigma_i$
are arbitrary (with $\sigma_2$ being negative and $\sigma_4$
positive). Note that one could absorb the factor of $6$ in
redefined parameters, but we prefer to keep it in order to
coincide with the notation of \cite{Sotiriou:2009bx}. As we
observe, the effect of the detailed-balance relaxation is the
decoupling of the coefficients, together with the appearance of a
term proportional to $a^{-6}$. In this case the corresponding
quantities for dark energy are generalized to
\begin{eqnarray}\label{rhoDEext}
&&\rho_{DE}|_{_\text{non-db}}\equiv
\frac{\sigma_1}{6}+\frac{\sigma_3 K^2}{6a^4} +\frac{\sigma_4
K}{6a^6}
\\
&&\label{pDEext} p_{DE}|_{_\text{non-db}}\equiv
-\frac{\sigma_1}{6}+\frac{\sigma_3 K^2}{18a^4} +\frac{\sigma_4
K}{6a^6}.
\end{eqnarray}
Again, it is easy to show that
\begin{eqnarray}\label{rhodotfluidnd}
\dot{\rho}_{DE}|_{_\text{non-db}}+3H(\rho_{DE}|_{_\text{non-db}}+p_{DE}|_{_\text{non-db}})=0.
\end{eqnarray}
 Finally, if we force (\ref{Fr1c}),(\ref{Fr2c}) to coincide with
 the standard Friedmann equations, we result to:
\begin{eqnarray}
&&G_{\rm cosmo}=\frac{6\sigma_0}{8\pi(3\lambda-1)}\label{simpleconstants0nd2}\\
&&\sigma_2=-3(3\lambda-1), \label{simpleconstants0nd}
\end{eqnarray}
while in this case the ``gravitational'' Newton's constant $G_{\rm
grav}$ reads \cite{Sotiriou:2009bx}:
\begin{eqnarray}
G_{\rm grav}=\frac{6\sigma_0}{16\pi}\label{Ggravbdb}.
\end{eqnarray}
Thus, the Friedmann equations take the standard form
(\ref{Fr1b}),(\ref{Fr2b}) too, but with $\rho_{DE}$ and $p_{DE}$
given by (\ref{rhoDEext}),(\ref{pDEext}) respectively.

\section{New agegraphic dark energy}
\label{NADEsec}

In this section we present the scenario of new agegraphic dark
energy. In order to be complete, we first construct the basic
model, and then we extend it in the case where the matter and dark
energy sectors interact with each other. Throughout the work, we
consider the background geometry to be Friedmann-Robertson-Walker.

\subsection{The basic scenario}

According to new agegraphic dark energy model \cite{Wei:2007ty}
the dark energy sector of the universe is attributed to a
holographic dark energy, where the IR cut-of of the theory is
taken to be the conformal time $\eta$ of the FRW universe:
\begin{equation}
\eta=\int\frac{dt}{a}=\int\limits_0^a\frac{da}{Ha^2}.
\end{equation}
Thus, the corresponding dark energy density reads:
\begin{equation}
\label{NADE} \rho_{DE}=\frac{3n^2}{8\pi G_{\rm grav}\eta^2},
\end{equation}
where the numerical factor $3n^2$ is introduced to parameterize
some uncertainties, such as the species of quantum fields in the
universe, the effect of curved spacetime (since the energy density
is derived for Minkowski spacetime), and so on \cite{Wei:2007ty}.

We stress here that, strictly speaking, the Newton's constant that
is present in (\ref{NADE}) is the gravitational one, since it
arises form the gravitational, black-hole properties of the
theory. As we discussed in the previous section, in conventional
theories this gravitational Newton's constant $G_{\rm grav}$
coincides with the cosmological one $G_{\rm cosmo}$, and their
distinction is not needed to be mentioned. However, since in the
present work we are interested in applying new agegraphic dark
energy in the framework of Ho\v{r}ava-Lifshitz gravity, in which
$G_{\rm grav}$ and $G_{\rm cosmo}$ do not coincide unless the IR
limit is reached, we prefer to maintain the distinction between
$G_{\rm grav}$ and $G_{\rm cosmo}$ in order to be transparent.

As usual, in the new agegraphic dark energy scenario, the energy
densities for matter and dark energy obey the standard evolution
equations:
\begin{eqnarray}
&&
\dot{\rho}_{M}+3H(\rho_{M}+p_{M})=0 \label{evoleq1a}\\
&&\dot{\rho}_{DE}+3H(\rho_{DE}+p_{DE})=0. \label{evoleq1}
\end{eqnarray}
In the following it proves more convenient to introduce the
density parameters
\begin{equation}
\Omega_M=\frac{8\pi G_{\rm cosmo}}{3H^2}\rho_M,\ \ \
\Omega_{DE}=\frac{8\pi G_{\rm cosmo}}{3H^2}\rho_{DE},\ \
\Omega_k=-\frac{k}{a^2H^2}. \label{densityparam}
\end{equation}
Thus, from (\ref{densityparam}) and (\ref{NADE}) we obtain
\begin{equation}
\label{OmDE}
 \Omega_{DE}=\left(\frac{G_{\rm cosmo}}{G_{\rm
grav}}\right)\frac{n^2}{H^2\eta^2}.
\end{equation}
Now, denoting by dot the time-derivative and by prime the
derivative with respect to $\ln a$, for every quantity F we
acquire $\dot{F}=F'H$. Thus, differentiating (\ref{OmDE}) and
using that $\dot{\eta}=1/a$  one obtains
\begin{equation}
\label{omegaDprim1e}
 \Omega'_{DE} =
-2\Omega_{DE}
\left[\frac{\dot{H}}{H^2}+\frac{\sqrt{\Omega_{DE}}}{na}\sqrt{\frac{G_{\rm
grav}}{G_{\rm cosmo}}}\right].
\end{equation}
Finally, differentiating (\ref{NADE}) we obtain
\begin{equation}
\label{rdeprime}
 \dot{\rho}_{DE}=-\frac{2\rho_{DE}}{a\eta},
\end{equation}
which allows us to use (\ref{evoleq1}) in order to define the new
agegraphic dark energy equation-of-state parameter as
\cite{Wei:2007ty,Wu:2008jt}:
\begin{equation}
\label{wde}
 w_{DE}=-1+\frac{2}{3na}\sqrt{\Omega_{DE}}\sqrt{\frac{G_{\rm
grav}}{G_{\rm cosmo}}}.
\end{equation}

\subsection{The interacting scenario}

A valuable extension of the aforementioned basic model, is the one
in which the matter and dark energy sectors are allowed to
interact \cite{interacting}, since such a scenario could alleviate
the known coincidence problem \cite{Steinhardt8}. In such a case
the evolution equations for the matter and dark energy densities
write:
\begin{eqnarray}
&&\dot\rho_M+3H(\rho_M+p_M)=Q\label{evoleq2a}\\
&&\dot\rho_{DE}+3H(\rho_{DE}+p_{DE})=-Q. \label{evoleq2}
\end{eqnarray}
A usual and quite general ansatz for the interaction term $Q$ is
$Q=3bH(\rho_M+\rho_{DE})$ \cite{Wei:2007ty}, with $b$ a coupling
parameter (we prefer to call the coupling $b$ and not $b^2$ in
order to coincide with the majority of the authors, since a
negative value could be possible too
\cite{Guo:2007zk,Feng:2008fx}). Therefore, $b<0$ corresponds to
energy transfer from dark matter to dark energy, while $b>0$
corresponds to dark energy transformation to dark matter. Finally,
in this case, relations (\ref{OmDE}), (\ref{omegaDprim1e}) and
(\ref{wde}) are valid too.

\section{New agegraphic dark energy in Ho\v{r}ava-Lifshitz gravity}
\label{NADEHL}

Let us now construct the scenario in which new agegraphic dark
energy is applied in a universe governed by Ho\v{r}ava-Lifshitz
gravity. That is, we will insert the relations (\ref{OmDE}) and
(\ref{omegaDprim1e}) of section \ref{NADEsec} in the Friedmann
equations derived of section \ref{model}. In order to be complete,
we perform this separately for the basic and for the interacting
case. Finally, note that we perform our analysis for a general
curvature, since, as it has been extensively stated in the
literature \cite{Calcagni:2009ar}, Ho\v{r}ava-Lifshitz cosmology
coincides completely with $\Lambda$CDM if one ignores curvature.

\subsection{The basic scenario}

Let us use the first Friedmann equation (\ref{Fr1b}) of
Ho\v{r}ava-Lifshitz cosmology, in order to eliminate the term
$\dot{H}/H^2$ that is present in new agegraphic dark energy
evolution (\ref{omegaDprim1e}). In order to simplify the
presentation we consider as usual that the matter is dust, that is
$w_M=0$. Differentiating (\ref{Fr1b}), using  (\ref{rdeprime}) and
(\ref{evoleq1a}), and inserting the density parameters
(\ref{densityparam}), we obtain
\begin{eqnarray}
\label{HdotH2}
 \frac{\dot{H}}{H^2} =-\Omega_k -\frac{3}{2}
\Omega_{M} -\frac{(\Omega_{DE})^{3/2}}{an}\sqrt{\frac{G_{\rm
grav}}{G_{\rm cosmo}}}.
\end{eqnarray}
Therefore, inserting this expression into  (\ref{omegaDprim1e}),
and using that $\Omega_{M}=1-\Omega_{DE}-\Omega_{k}$, we finally
acquire:
\begin{equation}
\label{omegaDprim1e2}
 \Omega'_{DE} =\Omega_{DE}\left[3(1-\Omega_{DE})-\Omega_k+ 2\sqrt{\frac{G_{\rm
grav}}{G_{\rm
cosmo}}}\frac{\sqrt{\Omega_{DE}}}{an}\left(\Omega_{DE}-1\right)\right].
\end{equation}
A final step is to use the definitions of $G_{\rm grav}$ and
$G_{\rm cosmo}$ in Ho\v{r}ava-Lifshitz cosmology. According to
(\ref{simpleconstants0a}) and (\ref{Ggrav}) the ratio $G_{\rm
grav}/G_{\rm cosmo}$ is equal to $(3\lambda-1)/2$ in the detailed
balance version of the theory, and it takes the same value in the
beyond detail-balance version too, as it can be extracted from
(\ref{simpleconstants0nd2}) and (\ref{Ggravbdb}). Therefore, in
both versions of Ho\v{r}ava-Lifshitz gravity, the differential
equation that determines the evolution of the new agegraphic dark
energy reads:
\begin{equation}
\label{omegaDprim1e3}
 \Omega'_{DE} =\Omega_{DE}\left[3(1-\Omega_{DE})-\Omega_k+
 2\sqrt{\frac{3\lambda-1}{2}}\frac{\sqrt{\Omega_{DE}}}{an}\left(\Omega_{DE}-1\right)\right].
\end{equation}

\subsection{The interacting scenario}

Let us repeat the above procedure in the case where the matter and
dark energy sectors are allowed to interact. Thus, differentiating
(\ref{Fr1b}), using (\ref{rdeprime}) and (\ref{evoleq2a}), we
obtain:
\begin{equation}
\label{HdotH2b}
 \frac{\dot{H}}{H^2} =-\Omega_k -\frac{3}{2}
\Omega_{M} -\frac{(\Omega_{DE})^{3/2}}{an}\sqrt{\frac{G_{\rm
grav}}{G_{\rm cosmo}}}+\frac{3}{2}b(\Omega_M+\Omega_{DE}).
\end{equation}
Therefore, inserting this relation into (\ref{omegaDprim1e}), and
using also that  $G_{\rm grav}/G_{\rm cosmo}=(3\lambda-1)/2$, we
acquire:
\begin{equation}
\label{omegaDprim1e4}
 \Omega'_{DE} =\Omega_{DE}\left[3(1-\Omega_{DE})-\Omega_k+
 2\sqrt{\frac{3\lambda-1}{2}}\frac{\sqrt{\Omega_{DE}}}{an}\left(\Omega_{DE}-1\right)-b(1-\Omega_k)\right].
\end{equation}
Finally, note that this relation holds in general, either in the
detailed-balance case, or beyond it.

\section{Cosmological implications}
\label{cosmimpl}

In the above sections we have formulated the new agegraphic dark
energy model in a universe governed by Ho\v{r}ava-Lifshitz
gravity, and in particular we extracted the differential equation
that determines the evolution of the dark energy density parameter
 $\Omega_{DE}$. Thus, in the present section we use this
 expression in order to calculate the basic observable, namely the
 dark energy equation-of-state parameter $w_{DE}$ at present as well as at
 small redshifts. In order to achieve this we perform the  standard
expansions of the literature. In particular, since $\rho_{DE}\sim
a^{-3(1+w_{DE})}$ we acquire \cite{Li:2004rb}
\begin{equation}
\ln\rho_{DE} =\ln \rho_{DE0}+{d\ln\rho_{DE} \over d\ln a} \ln a
+\frac{1}{2} {d^2\ln\rho_{DE} \over d(\ln a)^2}(\ln a)^2+\dots\ ,
\end{equation}
 where the
derivatives are calculated at the present time $a_0=1$ and the
index $0$ marks the value of a quantity at present. Then,
$w_{DE}(\ln a)$ is given as
\begin{equation}
w_{DE}(\ln a)=-1-{1\over 3}\left[{d\ln\rho_{DE} \over d\ln a}
+\frac{1}{2} {d^2\ln\rho_{DE} \over d(\ln a)^2}\ln a\right],
\end{equation}
 up to second order.
 Since
\begin{equation}
\rho_{DE} =\frac{3H^2\Omega_{DE}}{8\pi G_{\rm grav} }=\left(
\frac{G_{\rm cosmo}}{G_{\rm
grav}}\right)\frac{\rho_M}{\Omega_M}\Omega_{DE}=\left(
\frac{2}{3\lambda-1}\right)\frac{\rho_{m0}a^{-3}\Omega_{DE}}{(1-\Omega_k
-\Omega_{DE})},
\end{equation}
(which holds for dust matter) the derivatives in the
$w_{DE}$-expansion are easily computed using the obtained
expressions for $\Omega_{DE}'$. In addition, we can
straightforwardly calculate $w_{DE}(z)$, that is using the
redshift $z$ as the independent variable, replacing $\ln
a=-\ln(1+z)\simeq -z$, which is valid for small redshifts. Doing
so we obtain:
\begin{equation}
w_{DE}(z)=-1-{1\over 3}\left({d\ln\rho_{DE} \over d\ln a}\right)+
\frac{1}{6} \left[{d^2\ln\rho_{DE} \over d(\ln
a)^2}\right]\,z\equiv w_0+w_1z.
\end{equation}

\subsection{The basic scenario}

In this case  $\Omega_{DE}'$ is given by (\ref{omegaDprim1e3}) and
the aforementioned differentiation procedure leads to:
\begin{widetext}
\begin{equation}
\label{woni}
w_0=\frac{(1-\Omega_{k0})}{3n(1-\Omega_{k0}-\Omega_{DE0})}
\left\{n(\Omega_{k0}-3)+\sqrt{2(3\lambda-1)}\sqrt{\Omega_{DE0}}(1-\Omega_{DE0})+3n\Omega_{DE0}\right\}
\end{equation}
\begin{eqnarray}
\label{w1ni}
&&w_1=\frac{(\Omega_{k0}-1)\Omega_{DE0}}{12n\sqrt{3\lambda-1}\sqrt{\Omega_{DE0}}(-1+\Omega_{k0}+\Omega_{DE0})^2}
 \left\{\sqrt{2}n(3\lambda-1)(\Omega_{k0}-1)^2\right.\nonumber\\
 &&\ \ \ \ \ \ \ \left. +2n^2\sqrt{3\lambda-1}\sqrt{\Omega_{DE0}}\left[2(\Omega_{k0}-3)\Omega_{k0}\right]
\right.\nonumber\\
 &&\ \ \ \ \ \ \ \left.
 +2(\Omega_{k0}-1)(3\lambda-1)^{3/2}\sqrt{\Omega_{DE0}}+(3\lambda-1)\Omega_{DE0}^3\left(2\sqrt{3\lambda-1}\sqrt{\Omega_{DE0}}-3\sqrt{2}n\right)\right.\nonumber\\
 &&\ \ \ \ \ \ \ \left.+(3\lambda-1)\Omega_{DE0}^2
 \left\{6(\Omega_{k0}-1)\sqrt{3\lambda-1}\sqrt{\Omega_{DE0}}-\sqrt{2}n\left[14\Omega_{k0}-7\right]
 \right\}\right.\nonumber\\
 &&\ \ \ \ \ \ \ \left.+\Omega_{DE0}\left\{\sqrt{2}n(3\lambda-1)\left\{\Omega_{k0}\left[16-3\Omega_{k0}\right]-5
 \right\}\right.\right.\nonumber\\
 &&\ \ \ \ \ \ \ \ \ \ \ \ \ \ \ \ \ \ \ \ \  \ \ \ \ \ \left.\left.+12n^2\sqrt{3\lambda-1}\sqrt{\Omega_{DE0}}
 \Omega_{k0}-2(3\lambda-1)^{3/2}\sqrt{\Omega_{DE0}}(4\Omega_{k0}-3)
  \right\}  \right\}.
\end{eqnarray}
\end{widetext}

These expressions provide $w_0$ and $w_1$, for the basic scenario.
Despite their complicated form they can be very helpful since they
involve only the present density parameters
$\Omega_{k0}$,$\Omega_{DE0}$, the running parameter $\lambda$ of
Ho\v{r}ava-Lifshitz gravity, and the parameter $n$ of new
agegraphic dark energy. $\Omega_{k0}$,$\Omega_{DE0}$ are known in
a good accuracy, namely $\Omega_{DE0}=0.728_{-0.016}^{+0.015}$ and
$\Omega_{k0}=-0.013_{-0.007}^{+0.006}$  in 1$\sigma$
\cite{Komatsu:2010fb}. Additionally, observational constraints
restrict $\lambda$  in a narrow window around its IR value 1
namely $|\lambda-1|\leq 0.02$ in 1$\sigma$ \cite{Dutta:2010jh}.
Thus, the larger uncertainty comes from the value of the parameter
$n$, which varies in the bound $n=2.72^{+0.11}_{-0.11}$ in
1$\sigma$ \cite{Wei:2007xu}. Inserting these values and variation
intervals into (\ref{woni}),(\ref{w1ni}) we obtain that in
1$\sigma$:
\begin{eqnarray}
&&w_0=-0.780^{+0.022}_{-0.022}\nonumber\\
&&w_1=0.050^{+0.019}_{-0.018}.
\end{eqnarray}
As we observe, the value of the present dark-energy
equation-of-state parameter $w_0$ is slightly larger than the
standard new agegraphic dark energy model \cite{Wei:2007xu},
however, as expected, its 1$\sigma$-bounds are significantly
larger due to the additional uncertainties in the values of $n$
and $\lambda$. Finally, note that according to these values the
scenario at hand cannot exhibit the phantom-divide crossing. This
is a feature of the conventional new agegraphic dark energy model
\cite{Wei:2007ty,Wu:2008jt,Wei:2007xu} and it is inherited by the
Ho\v{r}ava-Lifshitz one too, since the increased bounds due to the
additional uncertainties are not too wide in order to cover the
values below $-1$.

\subsection{The interacting scenario}

For the interacting scenario,  $\Omega_{DE}'$ is given by
(\ref{omegaDprim1e4}) and repeating the above steps we result to:
\begin{widetext}
\begin{equation}
\label{woint}
w_0=\frac{(1-\Omega_{k0})}{3n(1-\Omega_{k0}-\Omega_{DE0})}
\left\{n\left[\Omega_{k0}-3+b(\Omega_{k0}-1)\right]+\sqrt{2(3\lambda-1)}\sqrt{\Omega_{DE0}}(1-\Omega_{DE0})+3n\Omega_{DE0}\right\}
\end{equation}
\begin{eqnarray}
\label{w1int}
&&w_1=\frac{(\Omega_{k0}-1)\Omega_{DE0}}{12n\sqrt{3\lambda-1}\sqrt{\Omega_{DE0}}(-1+\Omega_{k0}+\Omega_{DE0})^2}
 \left\{-\sqrt{2}n(3\lambda-1)(b-1)(\Omega_{k0}-1)^2\right.\nonumber\\
 &&\ \ \ \ \ \ \ \left. -2n^2\sqrt{3\lambda-1}\sqrt{\Omega_{DE0}}\left[b^2(\Omega_{k0}-1)^2-2(\Omega_{k0}-3)\Omega_{k0}+b(\Omega_{k0}-1)(3+\Omega_{k0})\right]
\right.\nonumber\\
 &&\ \ \ \ \ \ \ \left.
 +2(\Omega_{k0}-1)(3\lambda-1)^{3/2}\sqrt{\Omega_{DE0}}+(3\lambda-1)\Omega_{DE0}^3\left(2\sqrt{3\lambda-1}\sqrt{\Omega_{DE0}}-3\sqrt{2}n\right)\right.\nonumber\\
 &&\ \ \ \ \ \ \ \left.+(3\lambda-1)\Omega_{DE0}^2
 \left\{6(\Omega_{k0}-1)\sqrt{3\lambda-1}\sqrt{\Omega_{DE0}}-\sqrt{2}n\left[b(\Omega_{k0}-1)+14\Omega_{k0}-7\right]
 \right\}\right.\nonumber\\
 &&\ \ \ \ \ \ \ \left.+\Omega_{DE0}\left\{\sqrt{2}n(3\lambda-1)\left\{\Omega_{k0}\left[3b(\Omega_{k0}-1)-3\Omega_{k0}+16\right]-5
 \right\}\right.\right.\nonumber\\
 &&\ \ \ \ \ \  \ \ \ \ \ \left.\left.+6n^2\sqrt{3\lambda-1}\sqrt{\Omega_{DE0}}\left[b(\Omega_{k0}-1)+2\Omega_{k0}\right]-2(3\lambda-1)^{3/2}\sqrt{\Omega_{DE0}}(4\Omega_{k0}-3)
  \right\}  \right\}.
\end{eqnarray}
\end{widetext}
Similarly to the previous subsection, we can insert the values and
the 1$\sigma$-variation bounds for the model parameters
$\Omega_{k0}$,$\Omega_{DE0}$, $\lambda$, $n$ and the coupling
parameter $b$, in order to extract the corresponding values for
$w_0$ and $w_1$. For the first four we use
$\Omega_{DE0}=0.728_{-0.016}^{+0.015}$,
$\Omega_{k0}=-0.013_{-0.007}^{+0.006}$ \cite{Komatsu:2010fb},
 $|\lambda-1|\leq 0.02$
\cite{Dutta:2010jh} and $n=2.72^{+0.11}_{-0.11}$
\cite{Wei:2007xu}. For the coupling parameter $b$ there are
various observational constraints in the literature
\cite{Guo:2007zk,Feng:2008fx,Wang:2006qw}, all of which lie in a
narrow window around zero, thus we will use a representative
interval $-0.08<b<0.03$ \cite{Guo:2007zk}. Inserting these values
and variation intervals into (\ref{woint}),(\ref{w1int}), within
1$\sigma$ we obtain
\begin{eqnarray}
&&w_0=-0.82^{+0.08}_{-0.08}\nonumber\\
&&w_1=0.08^{+0.09}_{-0.07}.
\end{eqnarray}
As we observe, the value of $w_0$ is smaller than the
corresponding one of the non-interacting scenario above, while
$w_1$ is significantly larger. Furthermore, the bounds of both
these parameters are significantly larger, due to the additional
uncertainty in the coupling $b$. Finally, note moreover that these
bounds are also larger from the corresponding ones in interacting
new agegraphic dark energy models in conventional cosmology
\cite{Wei:2007xu}, due to the uncertainty in the running parameter
$\lambda$ of Ho\v{r}ava-Lifshitz  gravity. However, they are still
not so large in order to make the phantom-divide crossing
possible.

\section{Conclusions}
\label{Conclusions}

In this work we investigated the new agegraphic dark energy
scenario in a universe governed by Ho\v{r}ava-Lifshitz gravity. In
order to be general we considered both versions of the theory,
that is with or without the detailed-balance condition, we imposed
an arbitrary curvature for the background geometry, and we allowed
for an interaction between the matter and dark energy sectors. In
both the basic and interacting case we extracted the differential
equation that determines the evolution of the dark energy density
parameter, which is independent of the detailed-balance condition.
Finally, using this equation and performing a low-redshift
expansion of the dark energy equation-of-state parameter
$w(z)\approx w_0 +w_1 z$, we calculated $w_0$ and $w_1$ as
functions of the dark energy and curvature density parameters at
present, $\Omega_{DE0}$ and $\Omega_{k0}$ respectively, of the
running parameter $\lambda$ of Ho\v{r}ava-Lifshitz gravity, of the
parameter $n$ of new agegraphic dark energy, and of the
interaction coupling $b$.

In the non-interacting scenario, we found that
$w_0=-0.780^{+0.022}_{-0.022}$ is slightly larger comparing to the
standard new agegraphic dark energy model \cite{Wei:2007xu},
however its 1$\sigma$-bounds are significantly larger, as
expected, due to the additional uncertainties in the values of $n$
and $\lambda$. Furthermore, in the interacting case,
$w_0=-0.82^{+0.08}_{-0.08}$ is smaller than the corresponding one
of the non-interacting model, while $w_1=0.08^{+0.09}_{-0.07}$ is
significantly larger. Moreover, the bounds of both $w_0$ and $w_1$
are significantly larger, due to the additional uncertainty in the
interaction parameter $b$. Finally, note that these bounds are
also larger from the corresponding ones in interacting new
agegraphic dark energy models in standard cosmology
\cite{Wei:2007xu}, due to the uncertainty in the running parameter
$\lambda$ of Ho\v{r}ava-Lifshitz gravity. However, in both the
basic and interacting scenarios, the increased bounds in $w_0$ are
still not too large in order to make the phantom-divide crossing
possible.

It is interesting to note that the scenario of new agegraphic dark
energy in Ho\v{r}ava-Lifshitz cosmology seems to be more efficient
than that of holographic dark energy in the same gravitational
framework in a flat universe \cite{Setare:2010wt}. This feature
acts as an advantage of the present scenario, and indicates that
if the underlying gravitational theory is indeed the
Ho\v{r}ava-Lifshitz one and if dark energy exhibits a holographic
nature, then the new agegraphic version in a non-flat geometry
should be used instead of the simple (event-horizon) holographic
one.

We close this work by mentioning that although the present
analysis indicates that new agegraphic dark energy in
Ho\v{r}ava-Lifshitz cosmology can be consistent and compatible
with observations, it does not enlighten the discussion about
possible conceptual problems and instabilities of
Ho\v{r}ava-Lifshitz gravity, nor it can interfere with the
questions concerning the validity of its theoretical background,
which is the subject of interest of other studies. It just faces
the dark energy problem in such a context, and thus its results
can be taken into account only if Ho\v{r}ava-Lifshitz gravity
passes successfully the necessary theoretical tests.

\paragraph*{{\bf{Acknowledgements:}}}
ENS wishes to thank the Physics Department of Crete University for
the hospitality during the preparation of this work.

\end{document}